\documentclass[12pt, a4paper]{article}

\usepackage{amsmath,amssymb,bm,epsfig,afterpage}
\usepackage{cite}
\usepackage{here}
\usepackage{color}
\usepackage{colortbl}
\usepackage{xcolor}
\usepackage{tabularx}
\usepackage{subcaption}
\usepackage{bbm}
\usepackage{graphicx}
\usepackage{listings}
\usepackage{longtable}
\usepackage[utf8]{inputenc}
\usepackage{cancel}
\usepackage{slashed}

\makeatletter
    
    \@addtoreset{equation}{section}
  \makeatother

\allowdisplaybreaks[3]

\newcommand{\ka}{\kappa}

\newcommand{\kap}{{\kappa^\prime}}
\newcommand{\Zp}{{Z^\prime}}
\newcommand{\gp}{{g^\prime}}
\newcommand{\Qp}{Q^\prime}
\newcommand{\ve}{\mathbf{e}}

\newcommand{\vn}{\mathbf{n}}

\newcommand{\GeV}{\mathrm{GeV}}
\newcommand{\TeV}{\mathrm{TeV}}

\newcommand{\la}{\lambda}
\newcommand{\eps}{\epsilon}

\newcommand{\Mcal}{\mathcal{M}}
\newcommand{\Lcal}{\mathcal{L}}

\newcommand{\Ocal}{\mathcal{O}}
\newcommand{\Hcal}{\mathcal{H}}

\newcommand{\ev}{\mathbf{e}}

\newcommand{\nv}{\mathbf{n}}

\newcommand{\ol}[1]{\overline{#1}}

\newcommand{\vev}[1]{\langle{#1}\rangle}
\newcommand{\abs}[1]{\left|{#1}\right|}
\newcommand{\order}[1]{\mathcal{O}\left({#1}\right)}
\newcommand{\br}[2]{\mathrm{BR} \left({#1}\to{#2}\right)}
\newcommand{\dec}[2]{\Gamma \left({#1}\to{#2}\right)}

{\newcommand{\rep}[1]{\mathbf{#1}}

\newcommand{\eff}{\mathrm{eff}}

\newcommand{\bsll}{b\to s\ell\ell}
\newcommand{\damu}{\Delta a_\mu}
\newcommand{\Up}{U(1)^\prime}

\newcommand{\SRls}{\texttt{SR0}^{\mathrm{loose}}_{\mathrm{bveto}}}
\newcommand{\SRtt}{\texttt{SR0}^{\mathrm{tight}}_{\mathrm{bveto}}}
\newcommand{\SRfv}{\texttt{SR5L}}
\newcommand{\SRtl}{\texttt{SR2L}}
\newcommand{\os}{{\mathrm{OS}}}
\newcommand{\mTT}{m_{\mathrm{T2}}}
\newcommand{\MET}{E_T^{\mathrm{miss}}}
\newcommand{\SRZp}{\texttt{SRZp}}
\newcommand{\SRfp}{\texttt{SR5L}^\prime}

\newcommand{\disc}{{\mathrm{disc}}}
\newcommand{\excl}{{\mathrm{excl}}}

\usepackage[colorlinks=true, linkcolor=black, citecolor=black,
urlcolor=black]{hyperref}

\setcounter{footnote}{0}

\newif\iffigsame
\figsametrue

\begin{document}

\begin{titlepage}

\begin{flushright}
 {\tt
CTPU-PTC-21-14
}
\end{flushright}

\vspace{1.2cm}
\begin{center}
{\Large
{\bf
$\ge 4 \mu$ signal from a vector-like lepton decaying to a muon-philic $Z^\prime$ boson at the LHC
}
}
\vskip 2cm

Junichiro Kawamura$^{a,b}$~\footnote{jkawa@ibs.re.kr}
and
Stuart Raby$^c$~\footnote{raby.1@osu.edu}

\vskip 0.5cm

{\it $^a$
Center for Theoretical Physics of the Universe, Institute for Basic Science (IBS),
Daejeon 34051, Korea
}\\[3pt]

{\it $^b$Department of Physics,
Keio University, Yokohama 223-8522,  Japan} \\[3pt]

{\it $^c$
Department of Physics, Ohio State University, Columbus, Ohio, 43210, USA}\\[3pt]

\vskip 1.5cm

\begin{abstract}
We propose a novel possibility
to detect a very distinctive signal with more than four muons
originating from pair-produced vector-like leptons decaying
to a muon-philic $Z^\prime$ boson.
These new particles are good candidates to explain the anomalies
in the muon anomalous magnetic moment and the $b\to s\ell\ell$ processes. 
The doublet (singlet) vector-like leptons lighter than 1.3 (1.0) TeV
are excluded by the latest data at the LHC if $\mathrm{BR}(E\to Z^\prime \mu) = 1$.
We also show that the excess in the signal region with more than five leptons
can be explained by this scenario if the vector-like lepton
is a weak singlet, with mass about 400 GeV and $\mathrm{BR}(E\to Z^\prime \mu) = 0.25$.
The future prospects at the HL-LHC are discussed.
\end{abstract}

\end{center}
\end{titlepage}

\clearpage

\section{Introduction}

The Large Hadron Collider (LHC) explores
new physics beyond the Standard Model (SM) at TeV-scale.
The SM has been established as the theory just above the electroweak (EW) scale,
particularly by the discovery of the 125 GeV Higgs boson at the LHC~\cite{Aad:2012tfa,Chatrchyan:2012ufa}.
Although most of the experiments are consistent with the predictions of the SM,
there are 2-3 $\sigma$ discrepancies
in the measurements of rare semi-leptonic $B$ meson decays~\cite{Aaij:2014ora,Aaij:2017vbb,
Aaij:2019wad,Abdesselam:2019wac,Aaij:2013aln,
Lees:2013nxa,Aaij:2014pli,Aaij:2015esa,
Aaij:2013qta,Khachatryan:2015isa,Aaij:2015oid,Abdesselam:2016llu,Wehle:2016yoi,ATLAS:2017dlm,CMS:2017ivg,Aaij:2020ruw,Aaij:2021vac}, $\bsll$,
and the $4.2 \sigma$ discrepancy in
the anomalous magnetic moment of muon, $\Delta a_\mu$~\cite{Bennett:2006fi,Aoyama:2020ynm,Abi:2021gix}. 
An interesting coincidence here is that both anomalies are found
in physics related to muons, and hence these could be explained by the same origin.
One way to establish these discrepancies as evidence of new physics
is to increase the significance
by reducing the uncertainties in the experimental measurements
and in the predictions of the SM.
Another way is by directly discovering new particles at the LHC,
which we pursue in this paper.

It was shown in Refs.~\cite{Kawamura:2019rth,Kawamura:2019hxp}
that both anomalies in $\bsll$ and $\damu$ are addressed
by introducing vector-like (VL) fermions and a $\Zp$ boson associated
with an additional gauge symmetry $\Up$~\footnote{
See Refs.~\cite{Allanach:2015gkd,Altmannshofer:2016oaq,Megias:2017dzd,Raby:2017igl,King:2017anf,Darme:2018hqg} for models with VL fermions and $\Up$ for the anomalies.
The VL lepton explanation for $\damu$ is studied in e.g. Refs.~\cite{Czarnecki:2001pv,Kannike:2011ng,Dermisek:2013gta,Poh:2017tfo,Kawamura:2020qxo,Bai:2021bau}
}.
The former anomaly is explained by $\Zp$ exchange at the tree-level~\footnote{
The $\bsll$ anomaly can be explained by loop corrections involving VL families~\cite{Gripaios:2015gra,Arnan:2016cpy,Grinstein:2018fgb,Arnan:2019uhr,Chiang:2017zkh,Cline:2017qqu,Kawamura:2017ecz,Cerdeno:2019vpd,Arcadi:2021cwg}. 
},
while the latter is explained by loop corrections involving the VL leptons
and the $\Zp$ boson.
In this paper,
we point out the possibility that pair productions of VL leptons
provide very distinctive signals with more than four muons.
We shall discuss the current limits from the recent ATLAS data~\cite{Aad:2021qct}
and future prospects at the HL-LHC in a simplified model with a VL lepton
and $\Zp$ boson.
We also discuss the possible explanation for the excess
in the more than $5$-lepton signal found in Ref.~\cite{Aad:2021qct}.

The rest of this paper is organized as follows.
The simplified model is defined and then the relation to the anomalies are discussed
in Section~\ref{sec-simp}.
In Section~\ref{sec-LHC},
we discuss limits from the high-multiplicity lepton signal at the LHC.
Section~\ref{sec-Sum} is devoted to summary.
The model proposed in Refs.~\cite{Kawamura:2019rth,Kawamura:2019hxp}
are reviewed in Appendix~\ref{sec-VLmodel} as a UV completion of the simplified model.

\section{Simplified model}
\label{sec-simp}

We shall consider the simplified model with a VL lepton $E$,
which is weak singlet-like, $E_1$, or doublet-like, $L = (E_2, N)$,
where $N$ is the $SU(2)_L$ partner of $E_2$.
The $\Zp$ boson couplings to the leptons are given by
\begin{align}
\label{eq-simp}
 \Lcal_\Zp =&\  Z^\prime_\mu
\begin{pmatrix}
 \ol{\mu} & \ol{E}
\end{pmatrix}
\gamma^\mu
\left[
\begin{pmatrix}
 g^{L}_{\mu \mu} &  g^{L}_{\mu E} \\
 g^{L}_{\mu E}   &  g^{L}_{EE}    \\
\end{pmatrix}
  P_L  +
\begin{pmatrix}
 g^{R}_{\mu \mu} &  g^{R}_{\mu E} \\
 g^{R}_{\mu E}   &  g^{R}_{EE}    \\
\end{pmatrix}
P_R
\right]
\begin{pmatrix}
 {\mu} \\ {E}
\end{pmatrix} \\ \notag
&\ +
 Z^\prime_\mu
\begin{pmatrix}
 \ol{\nu} & \ol{N}
\end{pmatrix}
\gamma^\mu
\left[
\begin{pmatrix}
 g^{L}_{\nu \nu} &  g^{L}_{\nu N} \\
 g^{L}_{\nu N}   &  g^{L}_{NN}    \\
\end{pmatrix}
  P_L  +
\begin{pmatrix}
 0  &  0 \\
 0  &  g^{R}_{NN}    \\
\end{pmatrix}
P_R
\right]
\begin{pmatrix}
 {\nu} \\ {N}
\end{pmatrix},
\end{align}
where $E = E_1$ or $E_2$ and
the interactions with $N$ in the second line
are absent in the case of weak singlet VL lepton.
We assume that the off-diagonal couplings of the SM bosons
to the SM and VL leptons are negligible,
such that the dominant decay modes of the VL leptons are the decays
to a $\Zp$ boson and SM lepton.
In fact, this is achieved in the model proposed
in Refs.~\cite{Kawamura:2019rth,Kawamura:2019hxp}.

The loop corrections involving the VL leptons and the $\Zp$ boson
contribute to the anomalous magnetic moment of the muon.
It is known that the chiral-flip effect should be sizable to explain
the current discrepancy of $\order{10^{-9}}$ with the new particles above the EW scale.
In models with VL leptons,
the chiral-flip effects may come from the non-zero VEV of the SM Higgs doublet.
Hence the size of the loop correction is estimated as
\begin{align}
\label{eq-damuApp}
 \Delta a_\mu \sim&\ - \frac{m_\mu \ka v_H}{8\pi^2 m_\Zp^2} g^L_{\mu E} g^R_{\mu E}
                                C_{\damu}   \\  \notag
     \sim&\ 2.9\times 10^{-9} \times \left(\frac{500~\GeV}{m_\Zp}\right)^2
                          \left(\frac{\ka}{0.5}  \right)
                          \left(\frac{\sqrt{g^L_{\mu E} g^R_{\mu E}}}{0.25}\right)^2
                          \left(\frac{C_{\damu}}{0.1}\right),
\end{align}
where $\ka$ is the Yukawa couping constant for $\tilde{H}\ol{L}_R E_L$.
$C_{\damu}$ is the factor from loop functions which is typically of $\order{0.1}$,
see Appendix~\ref{sec-VLmodel} for the explicit form in the example model.

The $\Zp$ boson couplings to muons, $g_{\mu\mu}^L$ and $g_{\mu\mu}^R$,
directly relate to the Wilson coefficients for the $\bsll$ decay.
The effective Hamiltonian is given by~\cite{Buras:1994dj,Bobeth:1999mk}
\begin{align}
 \Hcal_\eff = - \frac{4 G_F}{\sqrt{2}} \frac{\alpha_e}{4\pi} V_{tb}V_{ts}^*
                       \left( C_9 \Ocal_9 + C_{10} \Ocal_{10}  \right),
\end{align}
where
\begin{align}
 \Ocal_9 := \bigl[\ol{s}\gamma^\mu P_L b \bigr]\bigl[\ol{\mu} \gamma_\mu \mu\bigr],
\quad
 \Ocal_{10} := \bigl[\ol{s}\gamma^\mu P_L b \bigr]\bigl[\ol{\mu} \gamma_\mu \gamma_5 \mu\bigr].
\end{align}
The coefficients in this model are given by
\begin{align}
  C_9 =&\  -\frac{\sqrt{2}}{4 G_F} \frac{4\pi}{\alpha_e}\frac{1}{V_{tb}V_{ts}^*}
   \frac{g^L_{sb}}{2m^2_{Z'}} \left(g^R_{\mu \mu} + g^L_{\mu\mu}\right) , \\
C_{10} =&\  -\frac{\sqrt{2}}{4 G_F} \frac{4\pi}{\alpha_e}\frac{1}{V_{tb}V_{ts}^*}
   \frac{  g^L_{sb}}{2m^2_{Z'}}\left(g^R_{\mu\mu} - g^L_{\mu\mu}\right),
\end{align}
where $g^L_{sb}$ is the $\Zp$ couplings to $\ol{s}b$ in the left-current.
The value of $C_9$ is estimated as
\begin{align}
\label{eq-C9app}
 \abs{C_9} \sim 0.87\times \left(\frac{m_\Zp}{500~\GeV} \right)^2
                             \left(\frac{g^L_{sb}}{0.0007} \right)
                             \left(\frac{g^L_{\mu\mu} + g^R_{\mu\mu}}{0.5} \right).
\end{align}
Note that the $\Zp$ couplings to quarks are tiny to explain the $\bsll$ anomaly,
while those to muons are large so that $\damu$ is explained
when $g_{\mu \mu}^{L,R} \sim g_{\mu E}^{L,R}$, which is true in the sample model.
This feature ensures that the $\Zp$ mass of $\order{100~\GeV}$ is not excluded
by the di-lepton resonance search at the LHC~\cite{Aad:2019fac}~\footnote{
See Refs.~\cite{Kohda:2018xbc,Allanach:2019mfl}
for general discussions for $\Zp$ boson responsible for $\bsll$
}.

In the $\Zp$ boson explanation, the ratio of the coefficients are given by
\begin{align}
 \frac{C_{10}}{C_{9}}
= \frac{g^{R}_{\mu\mu} - g^L_{\mu\mu}}{g^{R}_{\mu\mu} + g^L_{\mu\mu}}.
\end{align}
The recent analyses~\cite{Geng:2021nhg,Altmannshofer:2021qrr}
including the measurement of $R_K$ based
on the full run-2 data at the LHCb~\cite{Aaij:2021vac}~\footnote{
See Refs.~\cite{
Altmannshofer:2017fio,
Altmannshofer:2017yso,Alok:2017sui,Capdevila:2017bsm,Ciuchini:2017mik,DAmico:2017mtc,Geng:2017svp,Ghosh:2017ber,
Arbey:2018ics,Aebischer:2019mlg,Alguero:2019ptt,Alok:2019ufo,
Ciuchini:2019usw,Datta:2019zca,Kowalska:2019ley,
Arbey:2019duh,Kumar:2019nfv} for the analyses before the Moriond 2021.
}
favor $C_9$-only, $C_{10}$-only and $C_9=-C_{10}$ scenarios
among the one dimensional analyses,
which correspond to $g_{\mu\mu}^R = g_{\mu\mu}^L$,
$g_{\mu\mu}^R = - g_{\mu\mu}^L$ and $g^R_{\mu\mu} = 0$, respectively.
Among these three cases,
the explanation by $C_9=-C_{10}$ is not preferred
to explain the $\Delta a_\mu$ anomaly
because $\Delta a_\mu \propto g_{\mu\mu}^L g_{\mu\mu}^R$, see Eq.~\eqref{eq-damuApp}.
From this observation,
we shall consider the case with $\abs{g^L_{\mu\mu}} = \abs{g^R_{\mu\mu}}$
which predicts
\begin{align}
 \br{\Zp}{\mu\mu} := \frac{\dec{\Zp}{\mu\mu}}{\dec{\Zp}{\nu\nu} + \dec{\Zp}{\mu\mu}}
              \simeq \frac{ \abs{g^L_{\mu\mu}}^2 + \abs{g^R_{\mu\mu}}^2 }
                          {2\abs{g^L_{\mu\mu}}^2 + \abs{g^R_{\mu\mu}}^2 }
              = \frac{2}{3}.
\end{align}
Here,
we assume $g_{\mu\mu}^L = g_{\nu\nu}^L$ as expected from the $SU(2)_L$ symmetry
and
the $\Zp$ boson decay to quarks are negligible as expected from Eq.~\eqref{eq-C9app}.
Studies for $\abs{g^{L}_{\mu\mu}} \ne \abs{g^{R}_{\mu\mu}}$,
as preferred by the two dimensional analyses on $(C_9, C_{10})$ plane, are interesting 
but are beyond the scope of this paper.

\section{LHC signals}
\label{sec-LHC}

\begin{figure}[t]
\begin{minipage}[c]{0.32\hsize}
 \centering
\iffigsame
    \includegraphics[height=36mm]{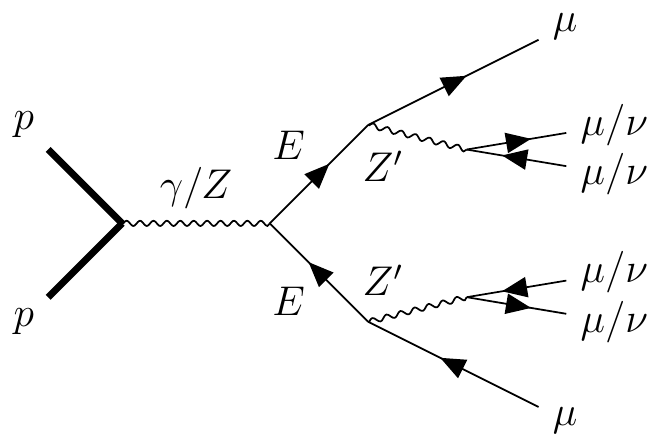}
\else
    \includegraphics[height=36mm]{graphics/tikz/prodEE.pdf}
\fi
\end{minipage}
\begin{minipage}[c]{0.32\hsize}
 \centering
\iffigsame
    \includegraphics[height=36mm]{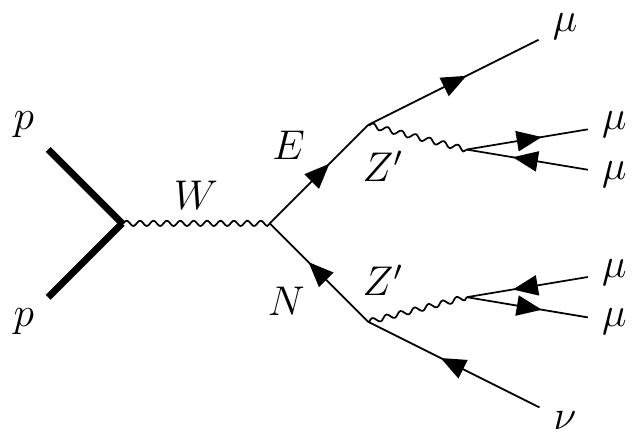}
\else
    \includegraphics[height=36mm]{graphics/tikz/prodNE.pdf}
\fi
\end{minipage}
\begin{minipage}[c]{0.32\hsize}
 \centering
\iffigsame
    \includegraphics[height=36mm]{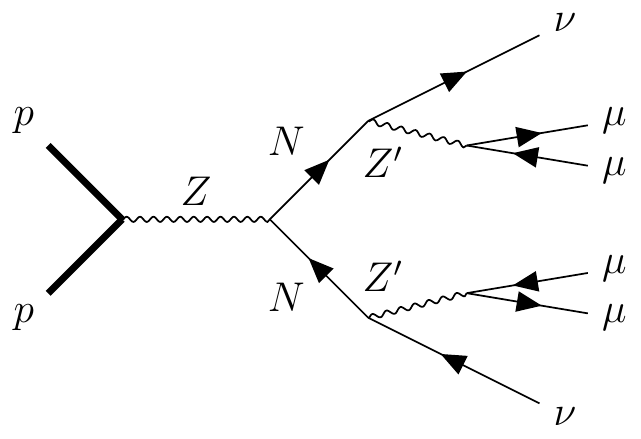}
\else
    \includegraphics[height=36mm]{graphics/tikz/prodNN.pdf}
\fi
\end{minipage}
 \caption{\label{fig-proc}
Processes produces more than four muons.
}
\end{figure}

In this paper, we study signals
from pair produced VL leptons decaying to the second generation leptons
and the $\Zp$ boson.
This can be realized when the VL leptons are heavier than the $\Zp$ boson.
If the $\Zp$ boson is heavier,
the VL lepton may decay to a SM boson and a lepton.
The limits for VL leptons in such a case are studied in Refs.~\cite{Aad:2015dha,Dermisek:2014qca,Sirunyan:2019ofn,Kumar:2015tna,Falkowski:2013jya,Ellis:2014dza,Bhattiprolu:2019vdu}.
It is also possible that the VL lepton decays to a new boson,
such as the physical mode of the $\Up$ breaking scalar.
Thus we treat the branching fraction of $E/N \to \Zp \mu/\nu$ as a free-parameter.
We further assume $\br{N}{\Zp \nu} = \br{E}{\Zp \mu}$ for simplicity.
Figure~\ref{fig-proc} shows the relevant processes
which can generate signals for more than four muons~\footnote{
We used TikZ-FeynHand to draw these figures~\cite{Ellis:2016jkw,Dohse:2018vqo}}.
Only the left process is relevant for the singlet-like case.

We recast the limits obtained in Refs.~\cite{Aad:2021qct} and~\cite{Aad:2019vnb}.
The former searches for signals with more than four leptons,
and the latter searches for signals with exactly two leptons 
with large missing transverse energy, $\MET$.
We have generated events using \texttt{MadGraph5$\_$2$\_$8$\_$2}~\cite{Alwall:2014hca}
based on a \texttt{UFO}~\cite{Degrande:2011ua} model file generated
with \texttt{FeynRules$\_$2$\_$3$\_$43}~\cite{Christensen:2008py,Alwall:2014hca}.
The events are showered with \texttt{PYTHIA8}~\cite{deFavereau:2013fsa}
and then run through the fast detector simulator \texttt{Delphes3.4.2}~\cite{deFavereau:2013fsa}.
We used the default ATLAS card for the detector simulation,
but the threshold on $p_T$ for the muon efficiency formula
is changed to $5$ GeV from $10$ GeV
since muons with $p_T>5~\GeV$ are counted as signal muons in Ref.~\cite{Aad:2021qct}.

\begin{table}[t]
 \centering
\caption{\label{tab-exrslt}
The number of events observed (data), fitted SM backgrounds (SM)
and  $95\%$ C.L. upper bound on the number of signal events $(S^{95})$
in the signal regions~\cite{Aad:2021qct,Aad:2019vnb}.
}
\begin{tabular}[t]{c|ccc|c} \hline
       & $\SRls$& $\SRtt$ & $\SRfv$ & $\SRtl$ \\  \hline \hline
data   & 11 & 1 & 21 & 37 \\
SM     & $11.5^{+2.9}_{-2.2}$ & $3.5^{+2.0}_{-2.2}$ & $12.4\pm 2.3$ & $37.3\pm 3.0$ \\
$S^{95}$ & $9.79$ & $3.87$ & $17.88$ & $14.3$ \\
\hline
\end{tabular}
\end{table}

We recast the experimental limits on the signal regions without
$Z$ boson, b-jet and hadronic $\tau$ defined in Ref.~\cite{Aad:2021qct}.\footnote{We recast
the analysis such that light leptons, ($e, \mu$), in Ref.~\cite{Aad:2021qct}, for our simple model these are just muons.}
These are named $\SRls$, $\SRtt$ and $\SRfv$.
The requirements for the events, in addition to the b-jet veto and hadronic $\tau$-veto,
common in the signal regions are as follows.
To meet the trigger thresholds,
$p_T$ of the leading muon, ordered by $p_T$, must be larger than $27~\GeV$,
or $p_T$'s of the leading and next-to leading muons are required
to be larger than $(15, 15)~\GeV$ or $(23, 9)~\GeV$.
If an opposite-sign (OS) muon pair whose invariant mass $m_\os$ is less than $4~\GeV$
or $8.4 < m_\os < 10.4~\GeV$, both leptons are discarded.
If two muons are found in $\Delta R < 0.6$ and one of them has $p_T < 30~\GeV$,
both leptons are discarded.
The first (second) $Z$ candidate is found from a pair of OS muons
whose $m_\os$ is the (second) closest to the $Z$ boson mass $m_Z = 91.2~\GeV$.
A pair is identified as a $Z$ boson if $m_\os \in [81.2, 101.2]$ $\GeV$.
Further, the event is considered to have a $Z$ boson if any system of
$\mu^+\mu^-\mu^\pm$ or $\mu^+\mu^-\mu^+\mu^-$ has invariant mass in $[81.1, 101.2]$ GeV.
In the signal regions $\SRls$ and $\SRtt$,
there must be more than four muons after the selections above,
and the event must not have any combinations of muons which is identified as a $Z$ boson.
Further, the effective mass of the event $m_\eff$,
defined as the scalar sum of $\MET$, $p_T$ of signal leptons
and $p_T$ of the jets with $p_T > 40$ GeV,
is required to be larger than 600 (1250)$~\GeV$ in the $\SRls$ ($\SRtt$).
In the $\SRfv$, the requirement is simply the lepton number to be larger than five,
and no further selection applied.

We also study the limits from the SUSY slepton search~\cite{Aad:2019vnb}
which requires exactly two leptons and large $\MET$.
The most relevant signal region for our scenario is
with same flavor (SF) two leptons without any jet.
There must be exactly two OSSF leptons, both with $p_T > 25~\GeV$.
Events are rejected if there are more muons with $p_T > 10$ GeV and $\abs{\eta} < 2.7$
or the two leading leptons are not opposite sign.
The missing energy $\MET$ and invariant mass of two leptons $m_\os$
must be larger than $110~\GeV$ and $121.2~\GeV$, respectively.
The stransverse mass $\mTT$~\cite{Lester:1999tx,Barr:2003rg} is required to be larger than $160$ GeV~\footnote{
We used the code provided by Ref.~\cite{Lester:2014yga} to calculate the stransverse mass.
}.
We name this signal region as $\SRtl$.

\begin{figure}[t]
 \centering
\iffigsame
    \includegraphics[height=0.5\hsize]{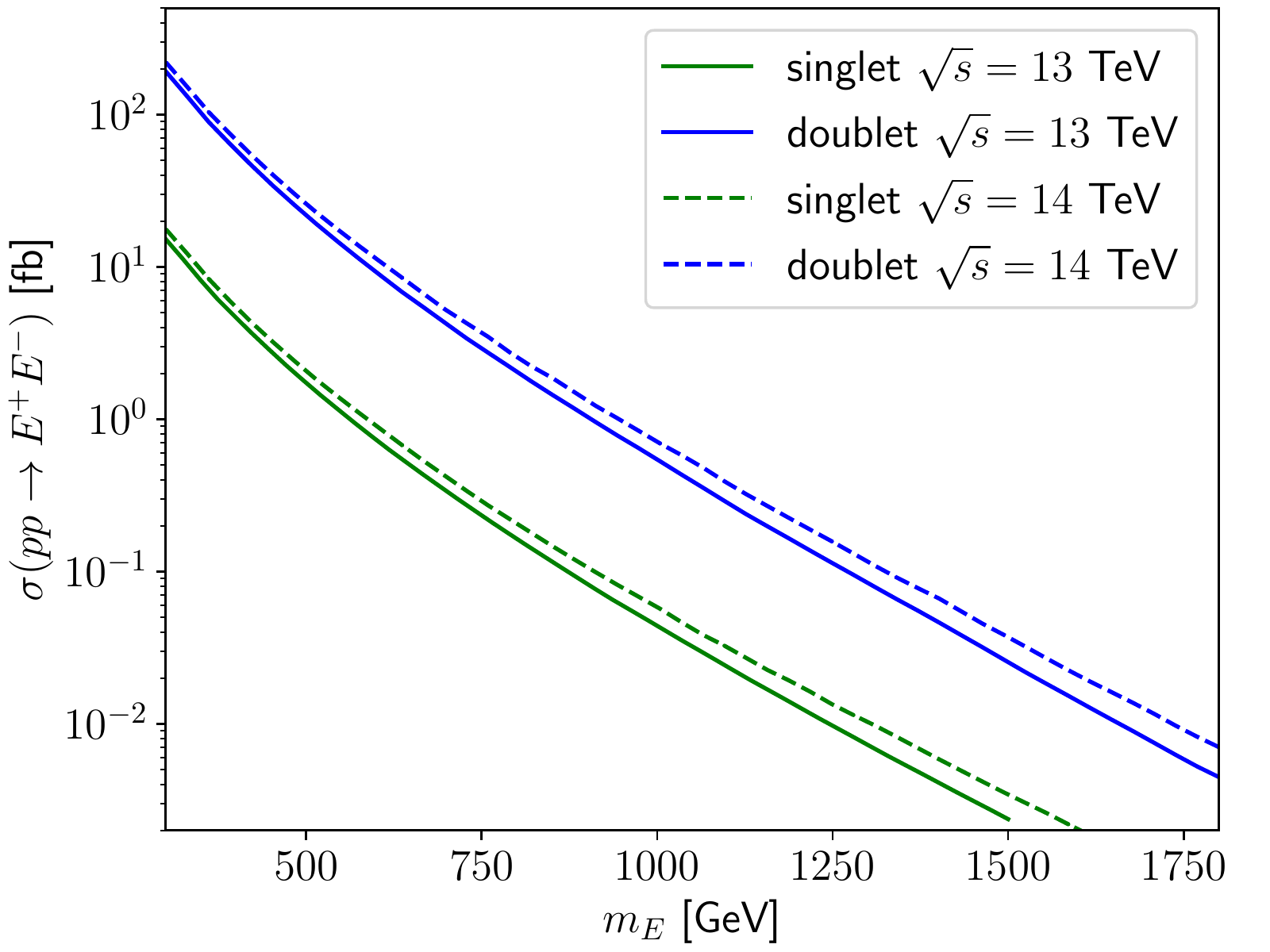}
\else
    \includegraphics[height=0.5\hsize]{graphics/fig_prod_0406.pdf}
\fi
 \caption{\label{fig-prod}
Pair-production cross section of VL leptons at $\sqrt{s}=13$ and 14 TeV.
The productions of VL neutrinos are included in the doublet case.
}
\end{figure}

The number of observed events, fitted SM backgrounds
and $95\%$ C.L. upper bounds on the signal events in the signal regions are shown
in Table~\ref{tab-exrslt}.
We see that there is an excess over the SM background in $\SRfv$
for which the local significance is $1.9\sigma$.
Figure~\ref{fig-prod} shows the production cross sections of VL leptons
at $\sqrt{s} = 13$ TeV and $14$ TeV calculated by \texttt{MadGraph5}.
We calculated the probability of how many events pass the cuts in each signal region
from the VL lepton pair production at $\sqrt{s} = 13$ TeV
by generating 25000 (50000) events at each point on the $(m_E, m_{\Zp})$ plane
for the singlet-like (doublet-like) VL leptons.

\subsection{Current limits}

\begin{figure}[h!t]
 \centering
\iffigsame
    \includegraphics[height=0.9\hsize]{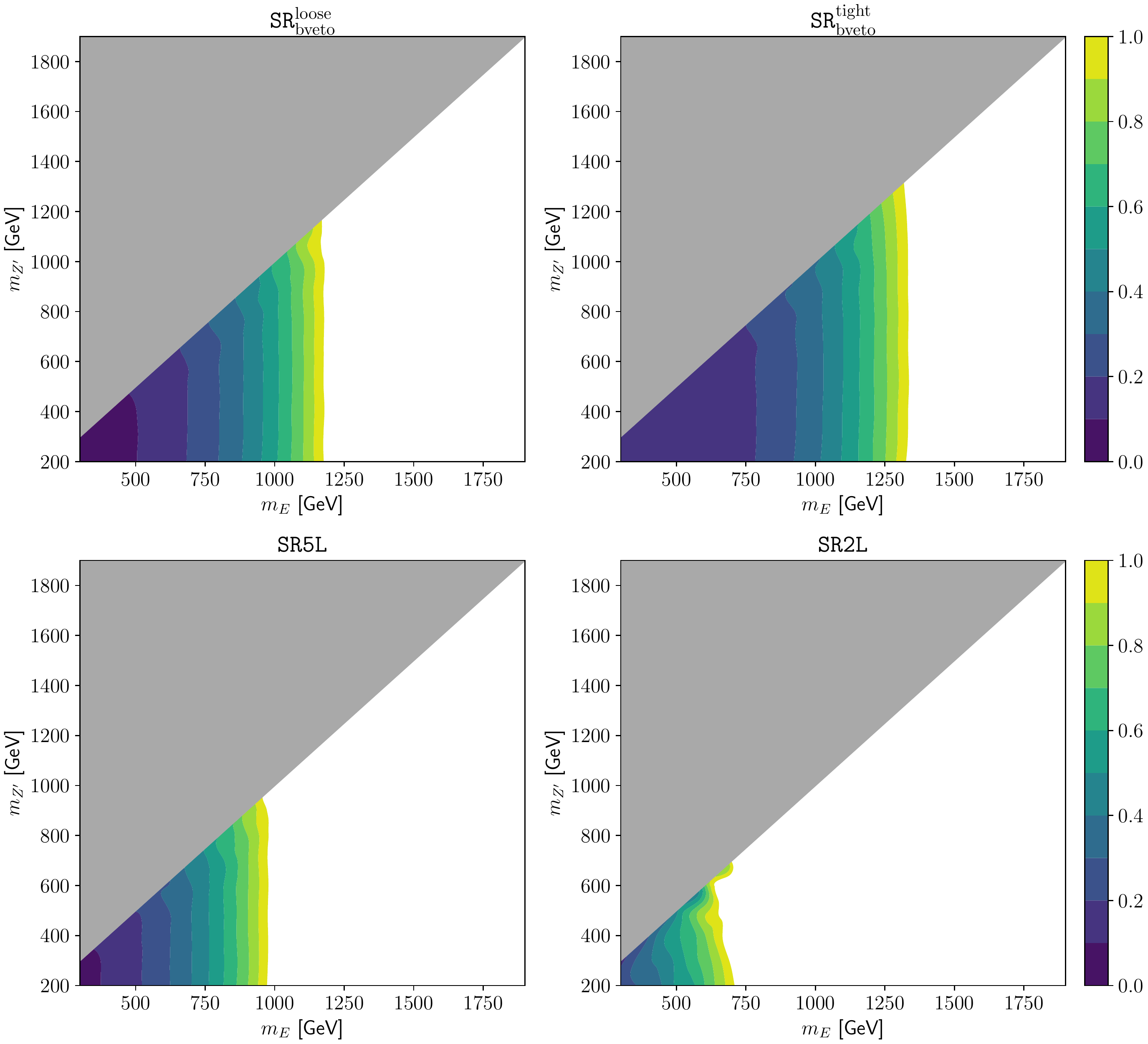}
\else
    \includegraphics[height=0.9\hsize]{graphics/fig_lims_doubletUC_0406.pdf}
\fi
 \caption{\label{fig-limD}
Limits on $\br{E}{\Zp \mu}$ for the doublet-like VL lepton.
}
\end{figure}

\begin{figure}[h!t]
 \centering
\iffigsame
    \includegraphics[height=0.45\hsize]{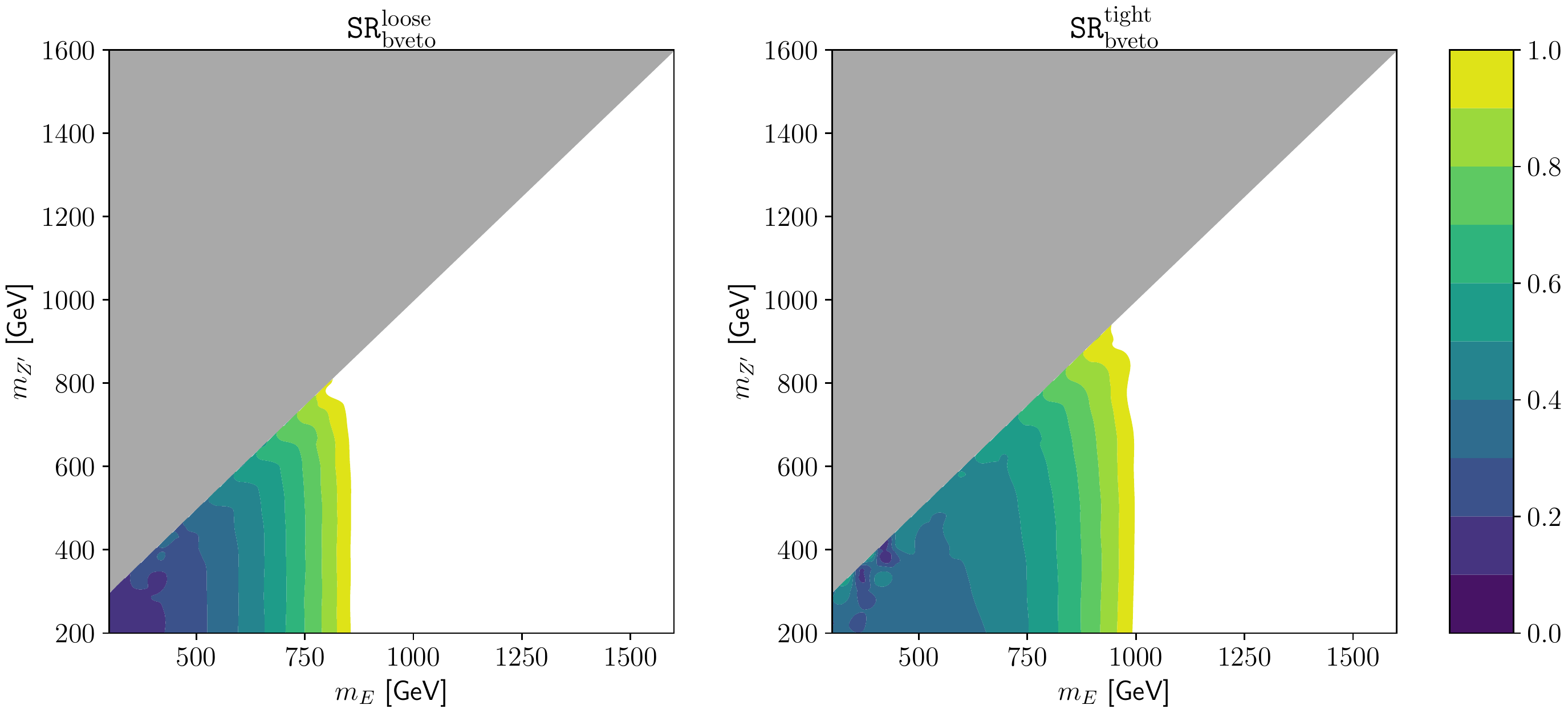}
\else
    \includegraphics[height=0.45\hsize]{graphics/fig_lims_singletUC_0406.pdf}
\fi
 \caption{\label{fig-limS}
Limits on $\br{E}{\Zp \mu}$ for the singlet-like VL lepton.
}
\end{figure}

Figures~\ref{fig-limD} and~\ref{fig-limS} show upper bounds on
$\br{E}{\Zp \mu}$ in the signal regions,
where the signal cross section is proportional to the branching fraction squared.
In the gray region, $m_\Zp > m_{E}$
and hence the decay $E \to \Zp \mu$ is kinematically forbidden.
The white region is not excluded by the current data
even if $\br{E}{\Zp\mu} = 1$.

For the doublet-like VL lepton,
we see that $\SRtt$ gives the strongest bound if $\br{E}{\Zp \mu} \gtrsim 0.2$,
because of fewer backgrounds satisfying the tighter $m_\eff$ cut.
The current limit is about 1350 GeV if $\br{E}{\Zp \mu} = 1$.
If the branching fraction is smaller, then $\SRls$ gives the stronger bound, since the cut by $m_\eff > 1250~\GeV$ of the $\SRtt$
is too tight for $m_{E_2} \lesssim 600$ GeV.
The limit from $\SRfv$ is weaker because of the excess
and that from $\SRtl$ is also weaker due to the larger backgrounds.
Note that the limits do not change much as the mass difference
between  the $\Zp$ boson and VL lepton decreases
since the signal muons can originate from $\Zp$ decays.

For the singlet-like VL lepton,
the strongest bound of about 1000 GeV is again from the $\SRtt$
if $\br{E}{\Zp \mu} = 1$.
The $\SRtt$ is not sensitive to the cases when $\br{E}{\Zp \mu} \lesssim 0.3$.
The difference comes from smaller production cross section of the singlet-like case.
The $\SRls$ gives the strongest constraint for smaller branching fractions.
The limits from $\SRfv$ and $\SRtl$ are not shown,
since the limits are much weaker than those from $\SRtt$ and $\SRls$
for the same reason as the doublet-like case.
In particular,  $\SRtl$  gives no bounds for the singlet-like case.

\subsection{Explanation for the excess in $\SRfv$}

\begin{figure}[!ht]
 \centering
\begin{minipage}[l]{0.48\hsize}
\iffigsame
    \includegraphics[height=0.75\hsize]{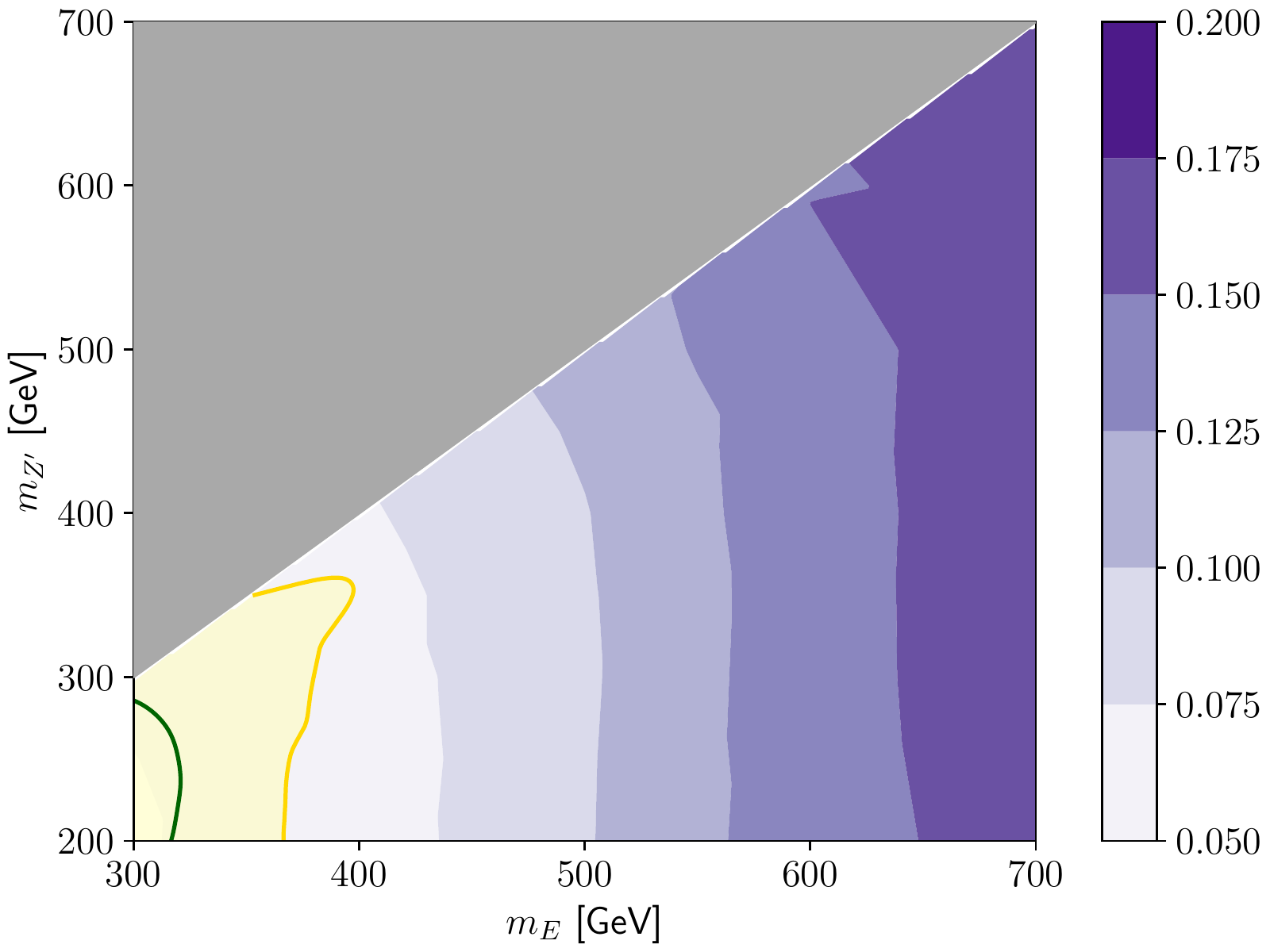}
\else
    \includegraphics[height=0.75\hsize]{graphics/fig_SR5lmax_doubletUC_0406.pdf}
\fi
\end{minipage}
\begin{minipage}[r]{0.48\hsize}
\iffigsame
    \includegraphics[height=0.75\hsize]{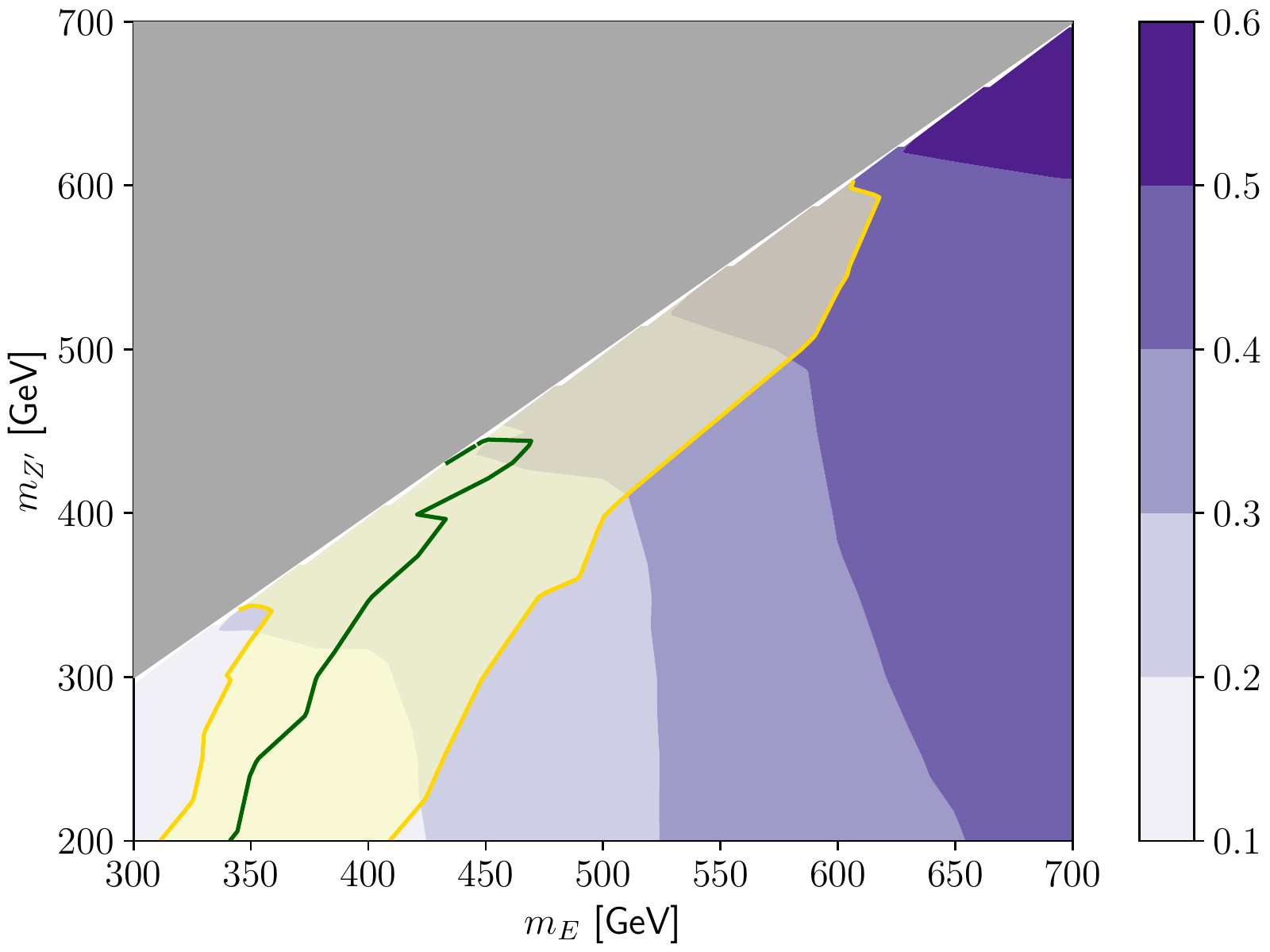}
\else
    \includegraphics[height=0.75\hsize]{graphics/fig_SR5lmax_singletUC_0406.pdf}
\fi
\end{minipage}
 \caption{\label{fig-S5lmax}
Maximum values of the number of signal events in $\SRfv$
consistent with the limits of $\SRls$ and $\SRtt$
for the doublet-like (singlet-like) VL lepton in the left (right) panel.
Background colors are the $\br{E}{\Zp\mu}$.
}
\end{figure}

\begin{figure}[!h]
 \centering
\begin{minipage}[l]{0.48\hsize}
\iffigsame
    \includegraphics[height=0.75\hsize]{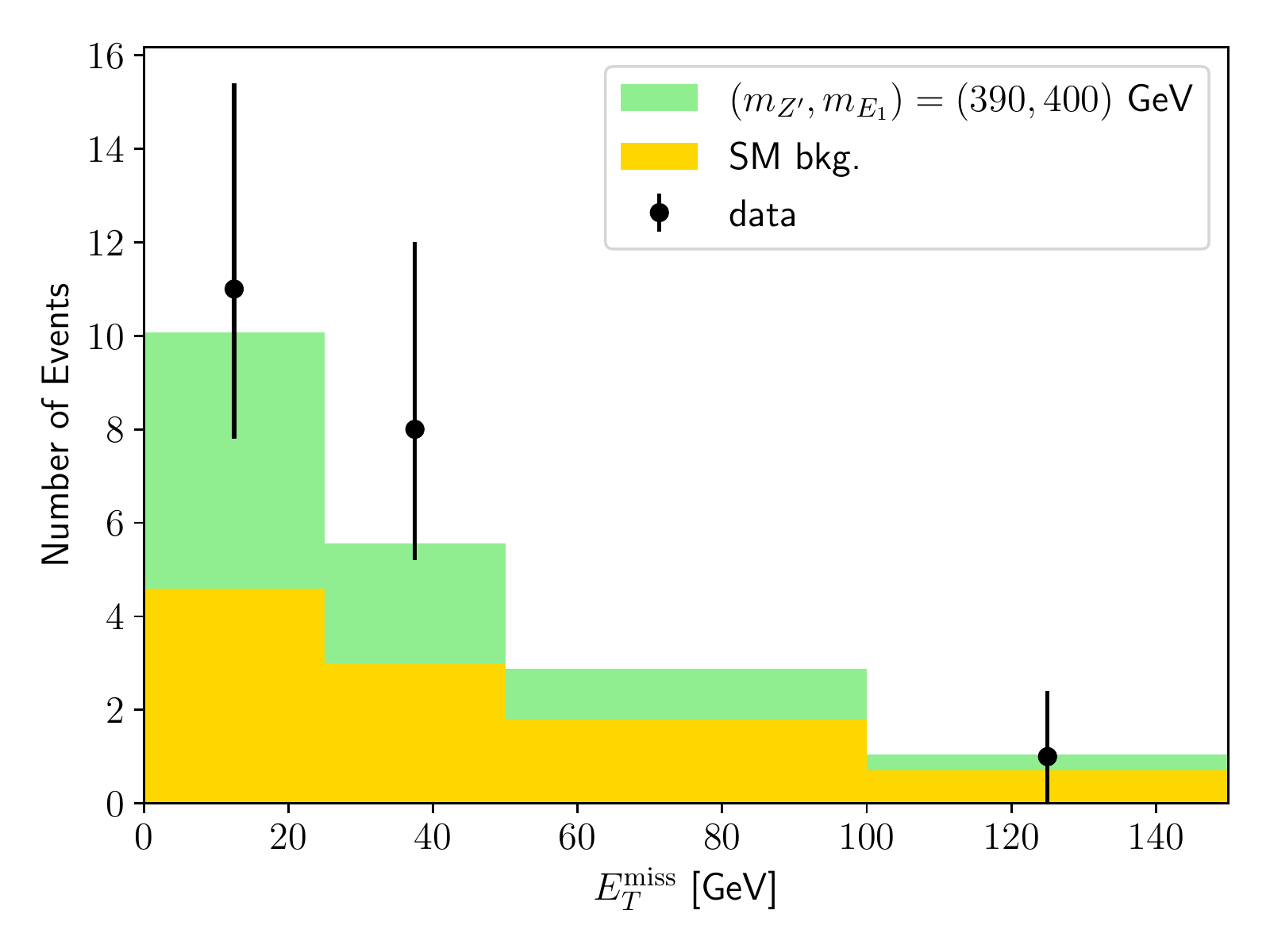}
\else
    \includegraphics[height=0.75\hsize]{graphics/fig_histMET_mZp390_mE400_mL2000_Br025.pdf}
\fi
\end{minipage}
\begin{minipage}[r]{0.48\hsize}
\iffigsame
    \includegraphics[height=0.75\hsize]{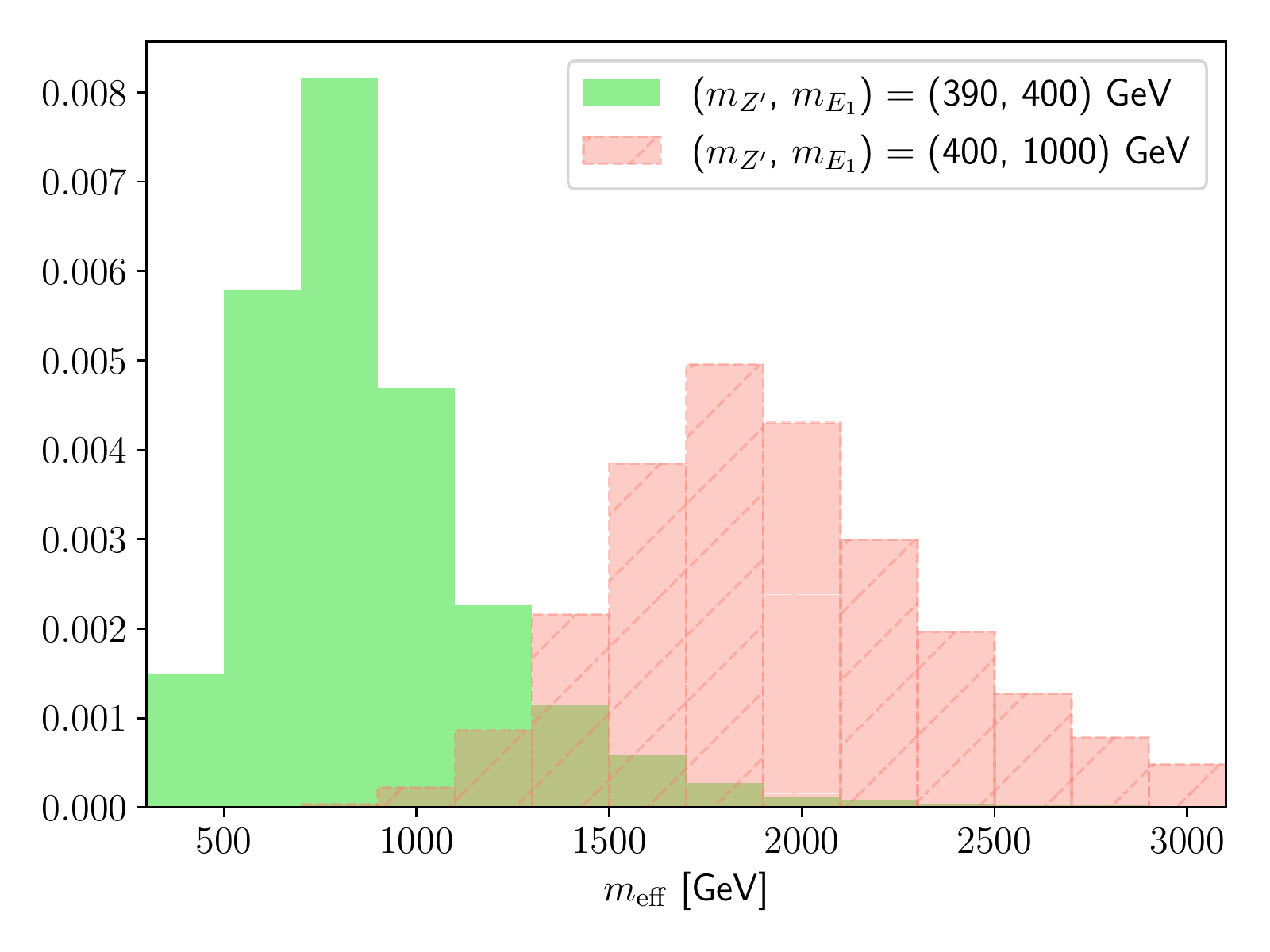}
\else
    \includegraphics[height=0.75\hsize]{graphics/fig_histMef_mZp390_mE400_mL2000_Br025.pdf}
\fi
\end{minipage}
 \caption{\label{fig-hists}
Histograms of $\MET$ (left) and $m_\eff$ (right).
In the left panel, $(m_\Zp, m_{E_1}) = (390, 400)~\GeV$ and $\br{E}{\Zp \mu} = 0.25$.
In the right panel, the values are normalized such that the total of the bins is unity.
}
\end{figure}

Figure~\ref{fig-S5lmax} shows the upper bound on the number of signal events
in the $\SRfv$ allowed by the limits from the other signal regions.
The background colors represent maximum values of $\br{E}{\Zp \mu}$.
Since the limits from the four lepton signal regions are severe,
the branching fraction should be so small that
the limits from $\SRtt$ and $\SRls$ are relaxed
by the fewer events passing the $m_\eff$ cut.
The excess is explained on the solid green line when the SM background is
at the central value shown in Table.~\ref{tab-exrslt}.
The yellow band corresponds to the uncertainty of the background estimation.
The singlet-like VL lepton can more easily explain the excess.
The limits from $\SRtt$ and $\SRls$ are much stronger for the doublet-like case,
since the production cross section is larger
and there are fewer muon signals originating from the VL neutrino production.

The left panel of Fig.~\ref{fig-hists} shows the $\MET$ distribution
after the selection of $\SRfv$ at
the benchmark point with $m_\Zp = 390~\GeV$, $m_{E_1} = 400~\GeV$
and $\br{E_1}{\Zp \mu} = 0.25$.
The SM contributions are represented by yellow bars
and the black dots (bars) show the data (its error bar),
which are read from Fig.~8 of Ref.~\cite{Aad:2021qct}.
We see that the $\MET$ distribution is well described by our scenario.
The benchmark point in our example model  
which realizes these masses and branching fractions 
and is also consistent with the muon anomalies at the same time 
is shown  Appendix~\ref{sec-bench}. 
The right panel of Fig.~\ref{fig-hists} shows the $m_\eff$ distribution
after the selection of the number of muons to be larger than four.
The values of the bars are normalized such that the sum of all the bins is unity.
The green bars are for the same benchmark point as the left panel,
and the red hatched bars are
for another benchmark point with $m_\Zp = 400~\GeV$ and $m_{E_1} = 1000~\GeV$.
We see that the peak of the distribution in $m_{\eff}$ is about $2m_{E_1}$,
and hence the strong constraint from $\SRtt$, which requires $m_\eff > 1250$ GeV,
is avoided and the tightest bound of $\br{E}{\Zp \mu} \lesssim 0.25$ is from $\SRls$.

We emphasize that this model can only explain the SR5L excess by muons.  
Thus the SR5L excess can not be explained, in this scenario,
if it includes signals with electrons.
The limits from the data would be significantly tightened
if lepton flavors are specified in the signal regions.
Thus the information of lepton flavor is crucial to test this model
with a muon-philic $\Zp$ and VL leptons.

The excess with electrons, might be explained
by VL leptons decaying to $Z$ or $W$ boson,
where the SM bosons decay leptonically.
If $m_\Zp > m_{E_1}$ which is the opposite case to our scenario,
the VL lepton will decay to a SM boson, including the Higgs boson.
In addition, electrons may come from the decays of the heavier VL leptons,
such as $E_2 \to E_1 Z$.
We note that the roughly degenerate mass of the VL leptons
are favored to explain the sizable $\Delta a_\mu$ in the model~\cite{Kawamura:2019hxp}.
These possibilities are interesting, but beyond the scope of this paper.

\subsection{Future prospects}
\label{sec-future}

\begin{figure}[!ht]
 \centering
\iffigsame
    \includegraphics[height=0.9\hsize]{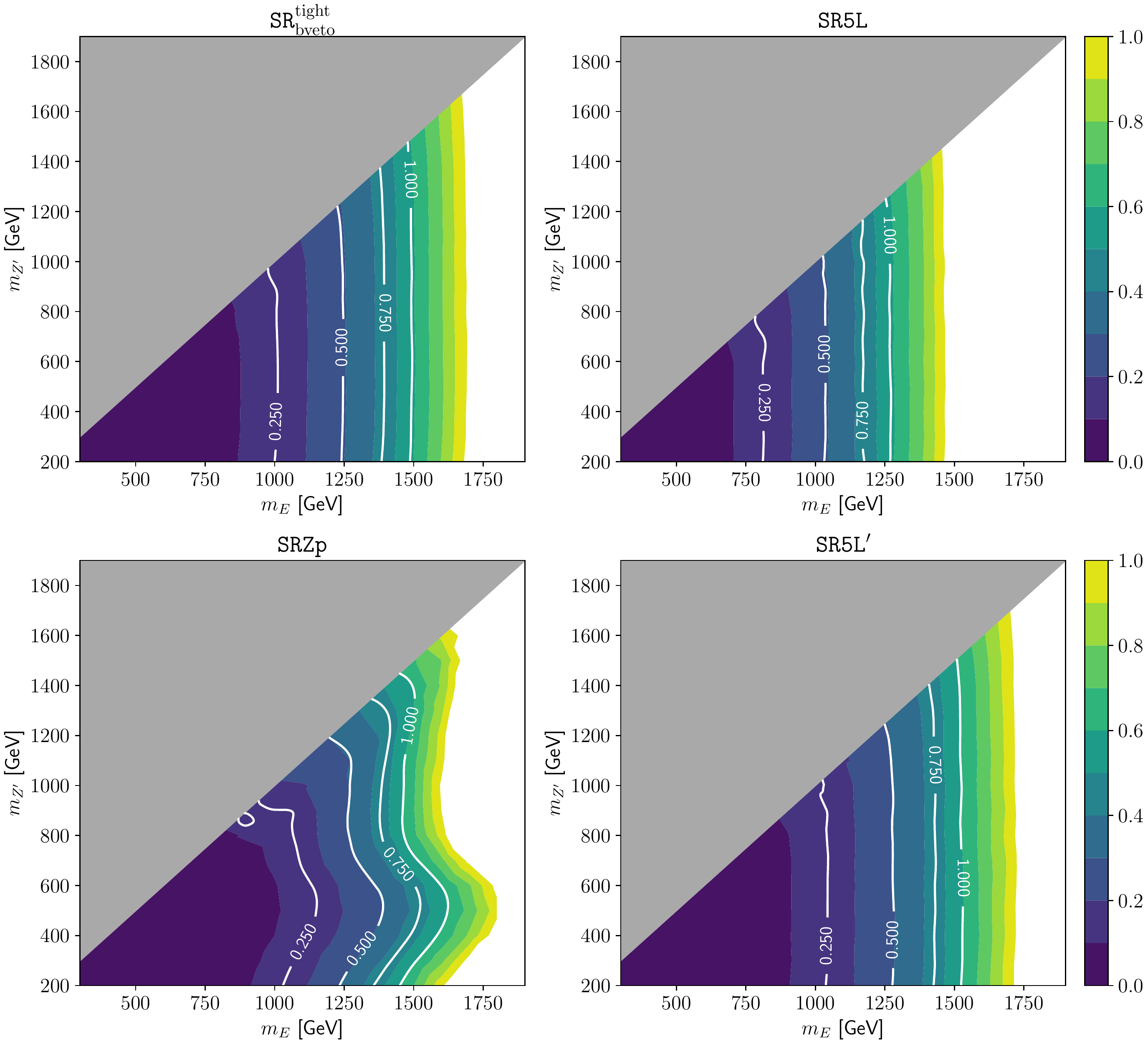}
\else
    \includegraphics[height=0.9\hsize]{graphics/fig_futurelims_doubletUC_0406.pdf}
\fi
 \caption{\label{fig-flimD}
Future prospects of the upper bound on $\br{E}{\Zp \mu}$
for the doublet-like VL lepton at the HL-LHC.
}
\end{figure}

\begin{figure}[!ht]
 \centering
\iffigsame
    \includegraphics[height=0.9\hsize]{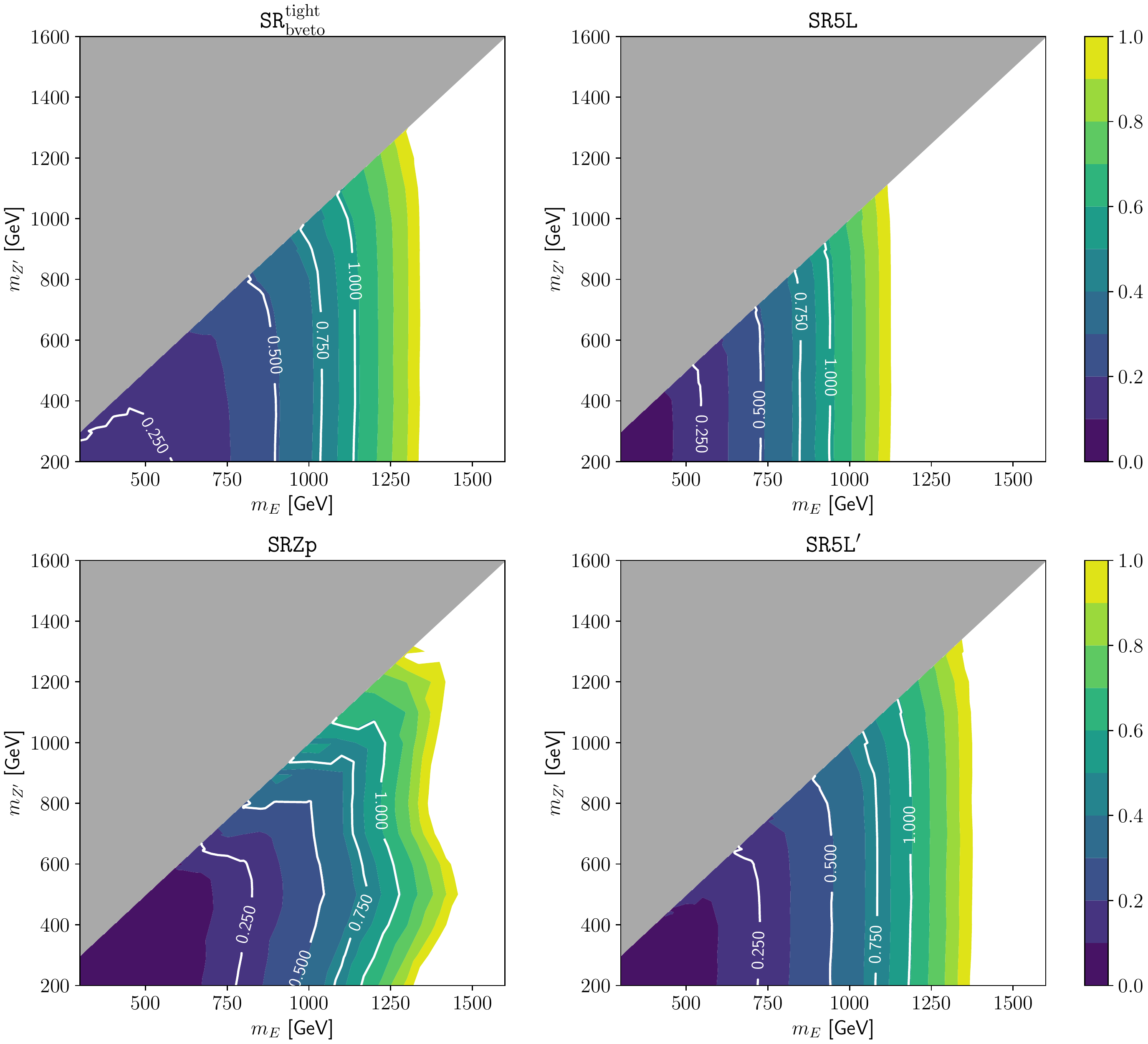}
\else
    \includegraphics[height=0.9\hsize]{graphics/fig_futurelims_singletUC_0406.pdf}
\fi
 \caption{\label{fig-flimS}
Future prospects of the upper bound on $\br{E}{\Zp \mu}$
for the singlet-like VL lepton at the HL-LHC.
}
\end{figure}

We shall discuss the discovery and exclusion potential at the HL-LHC
with $3~\mathrm{ab}^{-1}$ data.
We consider the two signal regions, $\SRtt$ and $\SRfv$
and propose two more new signal regions, $\SRZp$ and $\SRfp$.

We rescale the backgrounds to $\SRtt$ and $\SRfv$
by simply multiplying the ratio of integrated luminosity, $3000/139$.
For the signal events, we use the same efficiency times acceptance factor
as those used in the analyses for the current limits.
These are then multiplied by the integrated luminosity
and the production cross section at $\sqrt{s}=14~\TeV$
shown in Fig.~\ref{fig-prod} to calculate the number of signal events.

We define the two new signal regions, named $\SRZp$ and $\SRfp$.
In the $\SRZp$, at least four muons are required and the $Z$-veto is applied.
Then, two $\Zp$ candidates are chosen from any OS pair of muons,
where the reference $\Zp$ mass, $m_{\Zp}^{\mathrm{ref}}$, is set at $500~\GeV$,
in the same manner as the $Z$ candidates.
The (next-to) leading $\Zp$ candidate,
$m_\os$ is the (second) closest to $m_\Zp^{\mathrm{ref}}$,
must satisfy $\abs{m_\os - m_\Zp^{\mathrm{ref}}} < 100 $ ($250$) GeV.
We take relatively large range for the selection,
because we do not know the $\Zp$ mass.
The sensitivity can be improved by requiring strict range
for $\abs{m_\os-m_\Zp^{\mathrm{ref}}}$ and scan over $m_\Zp^{\mathrm{ref}}$
in the analysis, as in the $\Zp$ searches~\cite{Aad:2019fac,Sirunyan:2021khd}.
In the $\SRfp$, more than 5 muons are required.
Then, the $Z$-veto for any OS pair of muons and $m_\eff > 1000~\GeV$ cuts are applied.
In these two signal regions, we assume that there are 10 SM background events
per $3~\mathrm{ab}^{-1}$ data.
This may be a conservative assumption, since the cut is very tight
and not so many background events will survive,
c.f. the rescaled backgrounds in $\SRtt$ and $\SRfv$ are $76.6$ and $272$, respectively.

We quantify the future discovery and exclusion limits by the $p$-values
proposed in Ref.~\cite{Bhattiprolu:2020mwi},
\begin{align}
 p_\disc = \frac{\gamma(s+b,b)}{\Gamma(s+b)},
\quad
 p_\excl = \frac{\Gamma(b+1,s+b)}{\Gamma(b+1)},
\end{align}
where $s$ and $b$ are the number of signals and backgrounds.
$\Gamma(z)$, $\gamma(a,z)$ and $\Gamma(a,z)$ are the ordinary, lower incomplete
and upper complete Gamma functions.
The discovery (exclusion) limit corresponds to $p_\disc < 2.867\times 10^{-7}$
($p_\excl < 0.05$) where the significance is $>5$ ($>1.645$).
Here, we do not consider uncertainties in the signals and backgrounds for simplicity.

The future prospects at the HL-LHC for the doublet-like and the singlet-like
VL leptons are shown in Fig.~\ref{fig-flimD} and~\ref{fig-flimS}, respectively.
The background colors are the exclusion limits ($p_\excl < 0.05$)
for the branching fraction $\br{E}{\Zp \mu}$. 
The white lines show the discovery potential for a given branching fraction as labeled on the lines ($p_\disc < 2.867\times 10^{-7}$).
Assuming $\br{E}{\Zp\mu} = 1$, the doublet-like (singlet-like) VL lepton will be discovered
up to $m_E \lesssim 1.5$ ($1.15$) TeV by $\SRtt$. The limits from $\SRfv$ are weaker.
The $\SRZp$ may cover a wider parameter range than that of $\SRtt$
at $m_\Zp \sim 500~\GeV$ even if we set the relatively large range
for $\abs{m_\os-m_\Zp^{\mathrm{ref}}}$.
The $\SRfp$ may also cover a wider parameter range independent of the $\Zp$ mass,
based on our assumption of the background.
It is interesting that the entire parameter range with $\br{E}{\Zp \mu} \sim 0.25$,
which can explain the excess in $\SRfv$,
can be discovered in $\SRfv$, $\SRZp$ and $\SRfp$.

\section{Discussions}
\label{sec-Sum}

In this paper, we study the signal with more than four muons
originating from the pair-production of VL leptons decaying to a $\Zp$ boson
which couples to muons and/or muon neutrinos.
These particles may provide a way to resolve the tensions in the $\bsll$ decays and $\damu$.
The current limits can reach about $1~\TeV$ when the VL lepton decays
to the $\Zp$ boson exclusively owing to the very low backgrounds.
We showed that the excess in the signal region with five leptons or more 
may be explained in our model if the excess is given {\it only by muons}.
A benchmark point in our example model is given in Appendix~\ref{sec-bench} 
which simultaneously explains the muon anomalies as well as the excess 
in $\SRfv$.  
If the excess is not only muons,  
then the cascade decay of the heavier VL lepton
might be a nice candidate to explain the excess with electrons and muons in this kind of model.
The information of lepton flavor is crucial to test these new physics models
which explain the muon anomalies.

\vspace{0.5cm}
\textbf{Note added.}
While finalizing this manuscript,
the new experimental data of the muon anomalous magnetic moment was announced
from the FNAL.
The discrepancy from the SM prediction reaches to $4.2~\sigma$~\cite{Abi:2021gix}.
Models with VL leptons and a $\Zp$ boson coupling to muons
nicely explains the discrepancy of $\Delta a_\mu$, as well as the anomalies in $\bsll$.
These new particles could be confirmed by the LHC as discussed in this paper.

\section*{Acknowledgment}
The work of J.K.
is supported in part by
the Institute for Basic Science (IBS-R018-D1),
and the Grant-in-Aid for Scientific Research from the
Ministry of Education, Science, Sports and Culture (MEXT), Japan No.\ 18K13534.
The work of S.R. is supported in part by the Department of Energy (DOE) under Award No.\ DE-SC0011726.

\appendix

\section{Review of vector-like $U(1)^\prime$ model}
\label{sec-VLmodel}

\begin{table}[t]
\centering
 \caption{\label{tab-MTCT}
Matter contents.
Electric charge of fermion $f$ is $Q_f = T_f^3+Y_f/2$.
}
\begin{tabular}[t]{c|ccc|cccc|cc} \hline
              & $\ell_L$ & $\ol{\mu}_R$ & $H$ & $L_L$ & $\ol{E}_R$ & $\ol{L}_R$ & $E_L$
              & $Z^\prime$ & $\Phi$ \\ \hline \hline
$SU(2)_L$     & $\rep{2}$& $\rep{1}$& $\rep{2}$& $\rep{2}$&$\rep{1}$ & $\rep{2}$& $\rep{1}$ & $\rep{1}$ & $\rep{1}$ \\
$U(1)_Y$      & $-1$     & $2$      & $-1$     & $-1$      & $2$      & $1$ & $-2$ &$0$ & $0$\\
$U(1)^\prime$ & $0$      & $0$      & $0$      & $-1$      & $1$      &$1$ & $-1$
& $0$ & $-1$ \\
\hline
\end{tabular}
\end{table}

In this Appendix, we review the model proposed
in Refs.~\cite{Kawamura:2019rth,Kawamura:2019hxp}
as an example of a UV completion of the simplified model.
The matter contents of our model is given by Table~\ref{tab-MTCT}.
The $SU(2)_L$ doublets are defined as
\begin{align}
\ell_L = (\nu_L, \mu_L), \quad
 H =(H_0, H_-),\quad
 L_L = (N_L^\prime, E_L^\prime), \quad
 \ol{L}_R = (-\ol{E}_R^\prime, \ol{N}^\prime_R).
\end{align}
We only consider muons,
and assume that the couplings with the other leptons are negligible for simplicity.
The masses of VL states and Yukawa interactions are given by
\begin{align}
 \Lcal \supset&\ - m_L \ol{L}_R L_L -  m_E \ol{E}_R E_L  \\ \notag
         &\     + y_\mu \ol{\mu}_R \ell_L H
               + \kap \ol{E}_R L_L H  - \ka \ol{L}_R \tilde{H} E_L
                 + \la_L \Phi   \ol{L}_R \ell_L - \la_E \Phi \ol{\mu}_R E_L + h.c. ,
\end{align}
where $\tilde{H} := i\sigma_2 H^* = (H_-^*, -H_0^* )$.
The $SU(2)_L$ indices are contracted via $i\sigma_2$.
After the symmetry breaking by $v_H := \vev{H_0}$ and $v_\Phi := \vev{\Phi}$,
the mass matrix for the leptons are given by
\begin{align}
 \ol{\ve}_R \Mcal_e \ve_L  :=&\
\begin{pmatrix}
 \ol{\mu}_R & \ol{E}_R & \ol{E}_R^\prime
\end{pmatrix}
\begin{pmatrix}
 y_\mu v_H & 0        & \la_E v_\Phi \\
 0         & \kap v_H & m_E \\
\la_L v_\Phi & m_L & \ka v_H
\end{pmatrix}
\begin{pmatrix}
 \mu_L \\ E^\prime_L \\ E_L
\end{pmatrix},  \\
\ol{\vn}_R \Mcal_n \vn_L := &\
\ol{N}_R^\prime
\begin{pmatrix}
 \la_L v_\Phi & m_L
\end{pmatrix}
\begin{pmatrix}
 \nu_L \\ N_L^\prime
\end{pmatrix}.
\end{align}
The mass basis is defined as
\begin{align}
 \hat{\ve}_L := U_L^\dag \ve_L,
\quad
 \hat{\ve}_R := U_R^\dag \ve_R,
\quad
 \hat{\vn}_L := V_L^\dag \vn_L,
\quad
 \hat{\vn}_R := \vn_R,
\end{align}
where unitary matrices diagonalize the mass matrices as
\begin{align}
 U_R^\dag \Mcal_e U_L = \mathrm{diag}\left(m_\mu, m_{E_2}, m_{E_1}\right),
\quad
 \Mcal_n V_L =
\begin{pmatrix}
 0 & m_{N}
\end{pmatrix},
\end{align}
where $E_1$ ($E_2$) is the singlet-like (doublet-like) VL lepton~\footnote{
We restrict cases which $m_{E_1} \ll m_{E_2}$ or $m_{E_1} \gg m_{E_2}$,
so we can always identify the VL lepton is singlet-like or doublet-like.
}.
The non-zero mass of the SM neutrino will be explained
by introducing the right-handed counterparts,
but these are irrelevant for the present discussion.

We define the Dirac fermions as
\begin{align}
\ev := \left(\mu, E_2, E_1\right),  
\quad
\nv := \left(\nu, N\right),
\end{align}
 where
\begin{align}
\left[\ev\right]_i := \left( \left[\hat{\ev}_L\right]_i,
                          \left[\hat{\ev}_R \right]_i \right),
\quad
\nu := &\ \left( \left[\hat{\nv}_L \right]_1, 0 \right), \quad
 N   :=  \left( \left[\hat{\nv}_L \right]_2, N^\prime_R \right),
\end{align}
with $i=1,2,3$.

\subsection{Interactions}
The gauge interactions with the $\Zp$ boson in the mass basis are defined as
\begin{align}
 \Lcal_{V} =&\ \Zp_\mu \sum_{f=\ve, \vn}
                  \ol{f} \gamma^\mu \left({g}^\Zp_{f_L} P_L + {g}^\Zp_{f_R}P_R  \right)
                  f,
\end{align}
where the coupling matrices are given by
\begin{align}
 g^\Zp_{\ev_L} =&\ \gp U_L^\dag \Qp_e
  U_L, \quad
 g^\Zp_{\ev_R} = \gp U_R^\dag \Qp_e
  U_R,  \quad
g^\Zp_{\nv_L} = \gp V_L^\dag \Qp_n
V_L ,
\quad
g^\Zp_{\nv_R} = \gp  \Qp_n.
\end{align}
$P_{L}$ ($P_R$) are the chiral projections onto the left- (right-)handed fermions.
$\gp$ is the gauge coupling constant for $U(1)^\prime$.

We expand the neutral scalar fields as
\begin{align}
 H_0 = v_H + \frac{1}{\sqrt{2}} \left(h + i a_h\right),
\quad
 \Phi = v_\Phi + \frac{1}{\sqrt{2}} \left(\chi + i a_\chi\right),
\end{align}
where $h$ and $\chi$ are the physical real scalar fields,
while the pseudo-scalar components $a_h$ and $a_\chi$ are absorbed
by the $Z$ and $\Zp$ bosons, respectively.
The Yukawa interactions are given by
\begin{align}
-\Lcal_Y = \frac{1}{\sqrt{2}}
      \sum_{S=h, \chi}  \sum_{f=\ve,\vn} S \ol{f}\; Y^S_f P_L f + h.c. ,
\end{align}
where
\begin{align}
 Y^h_\ve = U_R^\dag
\begin{pmatrix}
y_\mu & 0 & 0 \\ 0 & \kap & 0 \\ 0 & 0 & \ka 		
\end{pmatrix}
U_L,
\quad
 Y^\chi_\ve = U_R^\dag
\begin{pmatrix}
0 & 0 & \la_E \\ 0 & 0 & 0 \\ \la_L & 0 & 0
\end{pmatrix}
U_L,
\quad
Y^h_\vn = 0_{2\times 2},
\quad
Y^\chi_\vn =
\begin{pmatrix}
 0 & 0 \\ \la_L & 0
\end{pmatrix}
V_L,
\end{align}

Let us define the approximate masses of the VL leptons as
\begin{align}
 M_L := \sqrt{m_L^2 + \la_L^2 v_\Phi^2},
\quad
 M_E := \sqrt{m_E^2 + \la_E^2 v_\Phi^2}.
\end{align}
Assuming $\ka v_H \ll \abs{M_L - M_E}$,
the diagonalization matrices for the charged lepton mass matrix are given by
\begin{align}
\label{eq-Uapp}
 U_L \sim
\begin{pmatrix}
 c_L & s_L & -\delta_L s_L \\ -s_L & c_L & -\delta_L c_L \\ 0 & \delta_L & 1
\end{pmatrix}
+ \order{\delta_L^2 }
,
\quad
 U_R \sim
\begin{pmatrix}
 c_R &  s_R \delta_R &  s_R \\ - s_R & c_R \delta_R & c_R \\ 0 & 1 &  - \delta_R
\end{pmatrix}
+ \order{\delta_R^2 },
\end{align}
where
\begin{align}
 c_L :=&\ \frac{m_L}{M_L},\quad s_L := \frac{\la_L v_\Phi}{M_L}, \quad
\delta_L := \frac{\ka v_H M_L}{M_L^2-M_E^2},  \\
 c_R :=&\ \frac{m_E}{M_E},\quad s_R := \frac{\la_E v_\Phi}{M_E}, \quad
\delta_R := \frac{\ka v_H M_E}{M_L^2-M_E^2}.
\end{align}
The diagonalized mass matrix is given by
\begin{align}
 U_R^\dag \Mcal_e U_L
 =
\begin{pmatrix}
\left(y_\mu c_L c_R + \kap s_L s_R \right) v_H  & \order{m_\mu} & 0 \\
0 & M_L + \order{\ka v_H \delta_{L,R}} & 0 \\
\order{m_\mu} & \order{\ka v_H \delta^2_{L,R}} & M_E + \order{\ka v_H \delta_{L,R}}
\end{pmatrix}
+ \order{m_\mu \delta_{L,R}},
\end{align}
where we assume $y_\mu v_H, \kap v_H \lesssim m_\mu$
to explain the muon mass without fine-tuning.

The $\Zp$ couplings in the mass basis are approximately given by
\begin{align}
\label{eq-gZpapp}
 g^\Zp_{\ev_L} =&\ -{\gp}
\begin{pmatrix}
 s_L^2  & -c_L s_L & c_L s_L \delta_L \\
 -c_L s_L & c_L^2 &  s_L^2 \delta_L \\
 c_L s_L \delta_L & s_L^2 \delta_L & 1
\end{pmatrix}
+ \order{\delta_L^2}, \\
\quad
 g^\Zp_{\ev_R} =&\ -{\gp}
\begin{pmatrix}
s_R^2 & -c_R s_R \delta_R & -c_R s_R \\
-c_R s_R\delta_R & 1 & -s_R^2 \delta_R \\
-c_R s_R & - s_R^2 \delta_R & c_R^2
\end{pmatrix}
+ \order{\delta_R^2}.
\end{align}
Hence, the effective couplings defined in Eq.~\eqref{eq-simp} are given by
\begin{align}
 \begin{pmatrix}
  g_{\mu\mu}^L & g_{\mu E}^{L} \\ g_{\mu E}^{L} & g_{EE}^L
 \end{pmatrix}
\sim -\gp
 \begin{pmatrix}
  s_L^2  & - s_L c_L \\ - s_L c_L & c_L^2
 \end{pmatrix},
\quad
 \begin{pmatrix}
  g_{\mu\mu}^R & g_{\mu E}^{R} \\ g_{\mu E}^{R} & g_{EE}^R
 \end{pmatrix}
\sim -\gp
 \begin{pmatrix}
  s_R^2  & - c_R s_R \delta_R \\ - c_R s_R \delta_R & 1
 \end{pmatrix},
\end{align}
in the doublet-like case, and these are given by
\begin{align}
 \begin{pmatrix}
  g_{\mu\mu}^L & g_{\mu E}^{L} \\ g_{\mu E}^{L} & g_{EE}^L
 \end{pmatrix}
\sim -\gp
 \begin{pmatrix}
  s_L^2  & c_L s_L \delta_L \\ c_L s_L \delta_L & 1
 \end{pmatrix},
\quad
 \begin{pmatrix}
  g_{\mu\mu}^R & g_{\mu E}^{R} \\ g_{\mu E}^{R} & g_{EE}^R
 \end{pmatrix}
\sim -\gp
 \begin{pmatrix}
  s_R^2  & - c_R s_R  \\ - c_R s_R & c_R^2
 \end{pmatrix},
\end{align}
in the singlet-like case.

The Yukawa couplings with $\chi$ are given by
\begin{align}
\label{eq-chiapp}
 Y^\chi_\ev \sim
\begin{pmatrix}
  0 & \la_E c_R \delta_L & \la_E c_R   \\
 \la_L c_L & \la_L s_L & \la_E s_R \delta_R - \la_L s_L \delta_L \\
 -\la_L c_L \delta_R & \la_E s_R\delta_L - \la_L s_L \delta_R & \la_E s_R
\end{pmatrix},
\quad
 Y^\chi_\nv \sim \la_L
\begin{pmatrix}
0 & 0 \\ c_L & s_L
\end{pmatrix},
\end{align}
where the coupling of $\chi$ with $\mu\mu$ is as small as $m_\mu/m_{E}$.
The couplings to the SM bosons are the SM-like up to $\order{m_\mu/m_{E}}$.

\subsection{Muon anomalies}

The $\Zp$ and $\chi$ boson contribution to $\Delta a_\mu$ is given by~\cite{Jegerlehner:2009ry,Dermisek:2013gta}
\begin{align}
\label{eq-delamu}
 \Delta a_\mu \sim
    \frac{  m_\mu \ka v_H}{64\pi^2 v_\Phi^2} s_{2L} s_{2R}  C_{LR},
\end{align}
with
\begin{align}
C_{LR} := \sqrt{x_L x_E} \frac{G_Z(x_L)-G_Z(x_E)}{x_L - x_E}
+ \frac{1}{2} \sqrt{y_L y_R} \frac{{y_L} G_{{S}}(y_L)-{y_R} G_{{S}}(y_R)}{y_L-y_R},
\end{align}
where $x_L := M_L^2/m_\Zp^2$, $x_E:= M_E^2/m_\Zp^2$,
$y_L:= M_L^2/m_\chi^2$ and $y_E := M_E^2/m_\chi^2$.
Here, $m_\Zp^2 = 2 \gp^2 v_\Phi^2$ is used.
The loop functions are given by
\begin{align}
  G_Z(x) :=&\  \frac{x^3+3x-6x \ln{(x)}-4}{2(1-x)^3}, \quad
  G_S(y) := \frac{y^2-4y+2\ln{(y)}+3}{(1-y)^3}.
\end{align}
The contribution from the scalar $\chi$ is included
since it is sizable unless $m_\chi$ is very heavy
which requires very large quartic couplings.

For the $b\to s\mu\mu$ anomaly, the Wilson coefficients are given by
\begin{align}
 C_9 \sim&\  - \frac{\sqrt{2}}{4G_F}\frac{4\pi}{\alpha_e} \frac{1}{V_{tb}V^*_{ts}}
                 \frac{1}{4v_\Phi^2} (s_R^2+ s_L^2 ) \eps_{Q_e} \eps_{Q_3}, \\
 C_{10} \sim&\  - \frac{\sqrt{2}}{4G_F}\frac{4\pi}{\alpha_e} \frac{1}{V_{tb}V^*_{ts}}
                 \frac{1}{4v_\Phi^2} (s_R^2- s_L^2 ) \eps_{Q_e} \eps_{Q_3},
\end{align}
where the $\Zp$ boson couplings to the SM doublet quarks are parametrized as
\begin{align}
 \left[g^{\Zp}_{d_L}\right]_{ij} \sim  \left[g^{\Zp}_{u_L}\right]_{ij} \sim
       -\gp \eps_{Q_i} \eps_{Q_j}.
\end{align}
$\eps_{Q_i}$ is the similar quantity as $s_{L} := \la_L v_\Phi/m_L$,
but we now consider the couplings with the second and third generation quarks
and these are typically small in contrast to that for muon.

From Eq.~\eqref{eq-delamu},
\begin{align}
 \Delta a_\mu \sim 2.9\times 10^{-9} \times
                 \left( \frac{1.0~\TeV}{v_\Phi}\right)^2
                 \left( \frac{\ka}{1.0} \right)
                 \left( \frac{s_{2L}s_{2R}}{1.0} \right)
                 \left( \frac{C_{LR}}{0.1} \right).
\end{align}
For the $b\to s\mu\mu$ anomaly,
\begin{align}
 C_9 \sim -0.62 \times \left(\frac{1.0~\TeV}{v_\Phi}\right)^2
                       \left(\frac{s_L^2+s_R^2}{1}\right)
                        \left(\frac{\eps_{Q_2}\eps_{Q_3}}{-0.002}\right).
\end{align}

Assuming $s_L = s_R = 1/\sqrt{2}$, i.e. $\la_L v_\Phi = m_L$ and $\la_E v_\Phi = m_E$,
the quark mixing angles are given by
\begin{align}
 \eps_{Q_2}\eps_{Q_3}
  \sim  -0.003 \times \left(\frac{C_9}{-0.82}\right)
                       \left(\frac{2.51\times 10^{-9}}{\Delta a_\mu}\right)
                       \left(\frac{\kappa}{1.0}\right)
                       \left(\frac{C_{LR}}{0.1}\right),
\end{align}
when the both anomalies are explained.
With such small couplings with quarks,
$\Zp$ boson is sufficiently suppressed to be consistent with the constraints
from the resonant di-lepton signal search at the LHC,
unless $\eps_{Q_2} \sim 1$ or $\eps_{Q_3} \sim 1$ to have large production cross section
from $s\ol{s}$ or $b\ol{b}$, respectively.

\subsection{Benchmark}
\label{sec-bench}

We show a benchmark scenario which explains the anomalies in $\damu$ and $\bsll$
and the excess in $\SRfv$ simultaneously.
As discussed in the main text,
singlet-like VL lepton is more suitable to explain the excess in $\SRfv$.
We take~\footnote{
With these values, $v_\Phi = 1103~\GeV$
which is sufficiently large to evade the bound from the neutrino trident process~\cite{Altmannshofer:2014cfa,Altmannshofer:2014pba,Magill:2016hgc,Ge:2017poy,Ballett:2018uuc,Altmannshofer:2019zhy}.
}
\begin{align}
m_\Zp =&\ 390~\GeV,  \quad M_L = 1.1~\TeV, \quad M_E = 404~\GeV,
\quad m_\chi = 365~\GeV, \\ \notag
s_L =&\ s_R = 1/\sqrt{2}, \quad y_\mu v_H = 2m_\mu,\quad
\kap = 0, \quad \ka = -0.821, \quad \gp = 0.25.
\end{align}
The VL lepton masses are $400$ and $1111~\GeV$.
The correction to the anomalous magnetic moment of the muon 
is $\Delta a_\mu = 2.51\times 10^{-9}$.
$s_L = s_R$ realizes the $C_9$-only scenario,
and $C_9 \sim -0.81$ is explained if $\eps_{Q_2} \eps_{Q_3} \sim -0.0032$.

The partial decay widths of the singlet VL lepton $E_1$ are approximately given by
\begin{align}
 \Gamma\left(E_1 \to \Zp \mu \right)
 \sim&\ \frac{M_E^3}{64\pi v_\Phi^2} c_R^2 s_R^2 (1-z)^2(1+2z),  \\
 \Gamma\left(E_1 \to \chi \mu \right)
 \sim&\ \frac{M_E^3}{64\pi v_\Phi^2} c_R^2 s_R^2 (1-x)^2,
\end{align}
where $z := m_\Zp^2/M_E^2$ and $x := m_\chi^2/M_E^2$.
Hence the branching fraction of $E_1$, assuming no other decay modes, is approximately given by
\begin{align}
\br{E_1}{\Zp \mu} \sim \frac{(1-z)^2 (1+2 z)}{(1-z)^2(1+2z) + (1-x)^2}.
\end{align}
At the benchmark point, $\br{E_1}{\Zp \mu} \simeq 0.25$.
The $\chi$ boson predominantly decays to VL fermions
as far as these are kinematically allowed.
If this is not the case, it should decay to a pair of SM leptons or quarks.
For the lepton coupling, as seen from Eq.~\eqref{eq-chiapp},
the coupling to 2 muons are strongly suppressed by the muon mass.
Hence, the dominant decay mode of $\chi$ may be to
a pair of top quarks due to the weaker suppression if $m_\chi > 2 m_t$
which is true at the benchmark point.
In this case,
the processes with $\chi$ decays will not contribute to the $\SRtt$ and $\SRls$
due to the b-jet veto, and thus the results in the main text will not be changed.
If $m_\chi < 2 m_t$, $\chi$ decays to a pair of bottom quarks,
where the relevant Yukawa is estimated as $\sim \eps_3 m_b/M_Q \sim 10^{-4}$
for $\eps_3 \sim 0.1$ and $M_Q \sim 4~\TeV$.
This would be comparable to the decay to a pair of muons
which the relevant Yukawa coupling is estimated as $\sim  m_\mu/M_L \sim 10^{-4}$
for $M_L \sim 1~\TeV$.
Thus, there will be additional contributions with six muons on top of the decays
from $\Zp$ boson.

\clearpage
{\small
\bibliographystyle{JHEP}
\bibliography{reference_vectorlike}

}

\end{document}